\documentstyle[twoside,fleqn,npb,epsfig]{article}
\def\S(#1){{{S}_{#1}}}
\def\Ss(#1,#2){{{S}_{#1,#2}}}
\def\Sss(#1,#2,#3){{{S}_{#1,#2,#3}}}
\def\Ssss(#1,#2,#3,#4){{{S}_{#1,#2,#3,#4}}}
\def\Sssss(#1,#2,#3,#4,#5){{{S}_{#1,#2,#3,#4,#5}}}

\newcommand{\beq}{\begin{equation}}
\newcommand{\eeq}{\end{equation}}
\newcommand{\bea}{\begin{eqnarray}}
\newcommand{\eea}{\end{eqnarray}}

\def\pqq(#1){p_{\rm{qq}}(#1)}
\def\Li(#1,#2){{\rm{Li}}_{#1}(#2)}
\def\H(#1,#2,#3,#4){{\rm{H}}_{#1,#2,#3}(#4)}

\newcommand{\AmS}{{\protect\the\textfont2
  A\kern-.1667em\lower.5ex\hbox{M}\kern-.125emS}}

\title{Tuning FORM with large calculations}

\author{J.A.M. Vermaseren\address{NIKHEF, P.O. Box 41882, \\ 1009 DB, 
Amsterdam}}

\begin{document}

\begin{abstract}
Some recent additions to FORM are discussed. In particular large file 
support and the tablebases are presented.
\end{abstract}

\maketitle

\section{Introduction}

Traditionally FORM~\cite{Vermaseren:2000nd} is called a program for 
particle theory. This is however a misconception that follows from a desire 
of putting labels on things. FORM is a program for many fields of science 
in which large formulae occur, like in deep perturbative expansions. Its 
dealing with non-commutative objects makes it also very suitable for 
mathematics calculations~\cite{jansa1,jansa2}. And it has also been used 
successfully in the field of Euler-Zagier sums~\cite{Vermaseren:1998uu} in 
which the results of certain categories of sums can only be obtained by 
solving large sets of equations. Of course FORM has been mostly tested in 
perturbative quantum field theory. However its speed and the potential size 
of its expressions should make FORM very attractive for many 
scientists.

New features are always looked at from a more generic viewpoint. This makes 
them useful for as many people as possible. Some of these new features are:
\begin{itemize}
\item
\$-variables which allow a high level of control over the organisation 
of a program. Version 1 and 2 of FORM never had this flexibility.
\vspace{-2mm} \item
Write to file facilities. This allows even dynamical addition to running 
programs.
\vspace{-2mm} \item
Large file support. Now also 32-bits processors can deal with intermediate 
expressions and files of more than 2 Gbytes, provided the operating system 
supports this (as in the ext-3 file-system versions of Linux).
\vspace{-2mm} \item
Better support for large tables.
\vspace{-2mm} \item
The tablebase. This is a database-like facility for extremely large tables. 
It was inspired by a calculation~\cite{Moch:2002xy,Vermaseren:2002xy} in 
which there were more than 20000 table elements, each of which occupied on 
average more than 20 Kbytes. To compile all these table elements in each 
program that might need some of the elements would be wasteful and slow, 
even in FORM. Now there are facilities by which the program can determine 
what is needed and when, and only those elements will be compiled at the 
proper moment.
\end{itemize}

\section{Some examples}

The first example concerns a run which is much like a benchmark used 
originally by D. Fliegner~\cite{Fliegner:1999jq} to test the parallel 
version of FORM. Later this test was taken over by R. 
Kreckel~\cite{kreckel} to compare the GiNaC system with other symbolic 
systems. Here we have modified it somewhat to allow the intermediate 
expression to surpass 4 Gbytes\footnote{A direct extension of the original 
test in which the power is 2 would run into the limit of 6000 variables (on 
32-bit systems) before the 4 Gbyte limit would be reached}.

{\footnotesize
\begin{verbatim}
FORM by J.Vermaseren,version 3.1(Jul 22 2002)
                  Run at: Mon Jul 22 15:06:39 2002
    #: SmallSize 10000000
    #: LargeSize 100000000
    #: TermsInSmall 1000000
    #define MAX "700"
    S	a0,...,a`MAX';
    L	F = (a0+...+a`MAX')^3;
    id	a1 = -a2-...-a`MAX';
    Print +f;
    .end

Time =    1.73 sec    Generated terms =     417057
             F      1 Terms left      =     133668
                      Bytes used      =    1882524
                          .
                          .
                          .
Time = 2611.92 sec    Generated terms =  457333100
             F      1 Terms left      =  390169111
                      Bytes used      = 5494312698

Time = 2613.52 sec
             F        Terms active    =  390169111
                      Bytes used      = 5496905398

Time = 2959.48 sec    Generated terms =  457333100
             F        Terms in output =          1
                      Bytes used      =         18
   F =
      a0^3;
\end{verbatim}
}
The run was on a notebook computer with a 850 MHz Pentium, 500 Mbytes 
memory and RedHat 7.3 Linux.

The next example shows the dynamic extension of tables during a run. It 
uses the \$-variables and the resulting table elements are also appended to 
a file. This way each new run can start by reading all results of the 
previous jobs. This mechanism was used to run and tabulate more than 20000 
integrals in the computation of basic building blocks for the three loop 
structure functions in deep inelastic scattering. Sometimes more than 1000 
integrals were done in a single (rather lengthy) run.

{\footnotesize
\begin{verbatim}
#include BE88fill.h
#do NUM = 1,500

  L FFK =
*   get an integral from a list
  #call intlist(BE88,`NUM')
        ;
*   for example this one:
*     +BE(0,1,1,1,1,2,1,3+N,0,1,0,0,0,0,0,N,0)
*
*  Here we compute the integral. Next we do
*
  L FFL =
*   get the integral again
  #call intlist(BE88,`NUM')
        ;
  id  BE(n1?,...,n7?,n8?!number_,k1?,k2?
                     ,0,0,0,0,0,N?!number_,k9?) =
      BE88fil(n1,...,n7,n8-N,k1,k2,k9,N)*f(be88);
* load arguments into $args and type into $ltype
  id  fx?{...,BE88fil,...}(?a$args,N)*f(x?$ltype)
                                             = 0;
  .sort
*   put the result in $expr
  #$expr = FFK;
*   'construct' a fill statement to add to table
  Fill `$ltype'fill(`$args') = `$expr';
  .global
*   make sure file is ready for appending
  #append <`$ltype'fill.h>
*   and append to file
  #write <`$ltype'fill.h> \
          "Fill `$ltype'fill(`$args') = %E;",FFK
  .store
#enddo
.end
\end{verbatim}
}

\section{The tablebase}

Faced with hundreds of megabytes of table elements of which we may 
typically need only a few in each job (but we cannot say in advance which) 
we need a special database structure. We want a database for FORM with the 
features:
\begin{itemize}
\item FORM reads at first only an index of the database.
\item At a specified time FORM can determine which elements are 
actually needed.
\item At a specified time FORM will load and compile these elements.
\item When the user speciefies it, the elements will be used.
\item The elements can be stored in gzipped~\cite{Gailly} form (saves a 
factor 4).\end{itemize}

Of course such `tablebases' need a number of control commands amoung which 
should be commands for
\begin{itemize}
\item Creating a new tablebase.
\item Adding tables and table elements to the tablebase.
\item Investigating what is in the tablebase.
\item Removing elements from the tablebase.
\item Cleaning up a tablebase.
\item Loading the index and compiling `stubbs'.
\item Loading and compiling individual elements.
\item Loading and compiling complete tables.
\item Loading and compiling indicated elements.
\item $\cdots$ and probably more $\cdots$.
\end{itemize}

The stubbs are intermediate expressions. They replace an object by an 
indicator that this table element exists in the tablebase. The advantage of 
this is that the object does not need to be manipulated by other routines 
that would deal with cases that are not in the tablebase, but yet we do not 
replace it by potentially lots of terms until we are ready for manipulating 
those terms.
Let us see how this works out.

{\footnotesize
\begin{verbatim}
 #-
 #define EXPANDEP "6"
 #include ensum.h
 #if `EXPANDEP' > 0
 S	ep(:`EXPANDEP');
 #endif
 .global
 L	F = x1+x2;
 .sort:start;
 #include be11fill.h
 #include be22fill.h
 #include be55fill.h
 #include be66fill.h
 #include be88fil1.h
 #include be88fil2.h
 #include be88fil3.h
 #include be88fil4.h
 #include la11fill.h
 #include la22fill.h
 #include la77fill.h
 #include no11fill.h
 #include no22fill.h
 .sort:after 4;
 .sort:complete reading;
 TableBase "three.tbl" create;
 .sort:create;
 TableBase "three.tbl" addto be11fill,be22fill
                   ,be55fill,be66fill,be88fill
                   ,la11fill,la22fill,la77fill
                   ,no11fill,no22fill;
 .sort:addto;
 .end
\end{verbatim}
}
This program gives the output
{\footnotesize
\begin{verbatim}
Time =    0.04 sec    Generated terms =          2
             F        Terms in output =          2
                start Bytes used      =         32

Time =  148.75 sec    Generated terms =          2
             F        Terms in output =          2
              after 4 Bytes used      =         32

Time =  148.76 sec    Generated terms =          2
             F        Terms in output =          2
     complete reading Bytes used      =         32

Time =  148.76 sec    Generated terms =          2
             F        Terms in output =          2
               create Bytes used      =         32
We add the name be11fill
We add the name be22fill
We add the name be55fill
We add the name be66fill
We add the name be88fill
We add the name la11fill
We add the name la22fill
We add the name la77fill
We add the name no11fill
We add the name no22fill

Time =  241.65 sec    Generated terms =          2
             F        Terms in output =          2
                addto Bytes used      =         32

Time =  241.65 sec    Generated terms =          2
             F        Terms in output =          2
                      Bytes used      =         32
\end{verbatim}
}
The running times refer to a Pentium 850. The first part shows the reading 
and compilation of the entire tables. The second part is the compression 
and the writing into the tablebase. How big are these files?

{\footnotesize
\begin{verbatim}
       lines     bytes
     
       21527   1665848 be11fill.h
       13123   1030971 be22fill.h
       12420    968211 be55fill.h
       19035   1486649 be66fill.h
      679908  53221903 be88fil1.h
      490372  38477216 be88fil2.h
      410549  32158987 be88fil3.h
      165495  12920526 be88fil4.h
      798355  61843593 la11fill.h
       37895   2896918 la22fill.h
      120615   9421657 la77fill.h
       48035   3647629 no11fill.h
       14177   1090916 no22fill.h
     2831506 220831024 total
     
      --->    51875476 three.tbl
     
      254795  20008483 ta0fill.h
       66790   5270950 ta1fill.h
      317318  24834908 ta2fill.h
      553843  43037497 ta3fill.h
      252002  19610471 ta5fill.h
      568589  44627190 tb0fill.h
      277903  21738353 tb1fill.h
       21553   1689338 tb5fill.h
     2312793 180817190 total
     
      --->    38599445 two.tbl
     
       35338   2348216 gtab00.prc
       49784   3313386 gtab01.prc
       48620   3065428 gtab10.prc
       68647   4383913 gtab11.prc
       51825   3283619 gtab20.prc
       51077   3401328 gtab02.prc
      305291  19795890 total
     
      --->     5005177 one.tbl
\end{verbatim}
}
These are the three loop, two loop and one loop tabulated integrals 
respectively.

There is already one pleasant spinoff. When we just load this one file 
three.tbl and enter and compile all elements we have

\noindent {\footnotesize
\begin{verbatim}
Time =    0.07 sec    Generated terms =          2
             F        Terms in output =          2
                start Bytes used      =         32

Time =    0.20 sec    Generated terms =          2
             F        Terms in output =          2
                 open Bytes used      =         32

Time =    0.59 sec    Generated terms =          2
             F        Terms in output =          2
                 load Bytes used      =         32

Time =  124.80 sec    Generated terms =          2
             F        Terms in output =          2
                enter Bytes used      =         32

Time =  124.80 sec    Generated terms =          2
             F        Terms in output =          2
                      Bytes used      =         32
\end{verbatim}
}
\noindent and we see that the fact that the file is compressed saves much 
time on the reading. Loading alone, the process of reading the index and 
compiling a complete list of `stubbs' takes about 0.4 sec. which indicates 
that we have eliminated the whole problem of slow startup.

Let us now try to use this. 
{\footnotesize
\begin{verbatim}
    #-
    .global
    L F = LA(1,N+1,1,1,1,1,1,1,0,N,0,0,0,0,0,0,3)
         +LA(1,N+1,1,8,1,1,1,1,0,N,0,0,0,0,0,0,3);
    Print +f +s;
    .sort

Time =    0.03 sec    Generated terms =          2
             F        Terms in output =          2
                      Bytes used      =        172

   F=
      +LA(1,1+N,1,1,1,1,1,1,0,N,0,0,0,0,0,0,3)
      +LA(1,1+N,1,8,1,1,1,1,0,N,0,0,0,0,0,0,3)
      ;
    TableBase "three.tbl" open;
    TableBase "three.tbl" load;
    .sort

Time =    0.40 sec    Generated terms =          2
             F        Terms in output =          2
                      Bytes used      =        172
   id  LA(n1?pos_,n2?!number_,<n3?pos_>,...,
          <n8?pos_>,k1?,k2?!number_,k3?,0,0,0,0,
             0,k9?) = LA22(n1,...,n8,k1,k2,k3,k9);
   Print +f +s;
   .sort

  F=
     +LA22(1,1+N,1,1,1,1,1,1,0,N,0,3)
     +LA22(1,1+N,1,8,1,1,1,1,0,N,0,3)
     ;
   *
   * Shift to table notation and back
   * Whatever is in the table will be intercepted
   *
   id  LA22(n1?,...,n8?,k1?,k2?,k3?,k9?) =
         la22fill(n1,n2-k2,n3,...,n8,k1,k3,k9,k2);
   id  la22fill(n1?,...,n8?,k1?,k3?,k9?,k2?) =
           LA22(n1,n2+k2,n3,...,n8,k1,k2,k3,k9);
   Print +f +s;
   .sort

  F=
     +tbl_(la22fill,1,1,1,1,1,1,1,1,0,0,3,N)
     +LA22(1,1+N,1,8,1,1,1,1,0,N,0,3)
     ;
   TestUse la22fill;
   Print +f +s;
   .sort

  F=
     +tbl_(la22fill,1,1,1,1,1,1,1,1,0,0,3,N)
     +LA22(1,1+N,1,8,1,1,1,1,0,N,0,3)
     ;
   TableBase "three.tbl" use;
   Print +f +s;
   .sort

Time =    0.45 sec    Generated terms =          2
             F        Terms in output =          2
                      Bytes used      =        152

  F=
     +tbl_(la22fill,1,1,1,1,1,1,1,1,0,0,3,N)
     +LA22(1,1+N,1,8,1,1,1,1,0,N,0,3)
     ;
   PolyFun acc;
   Apply;
   id Nval(N?)*R(n?,x?) = den(x+N)^n;
   id Nval(N?)*R(n?,x?,?a) =
                          den(x+N)^n*S(R(?a),x+N);
   id S(R,x?) = 1;
   id Nval(N?) = 1;
   id  z3?{z3,z4,z5,z6} = acc(z3);
   id  ep^n? = acc(ep^n);
   Print +f +s;
   B   theta,delta;
   .end

Time =    0.45 sec    Generated terms =         50
             F        Terms in output =         32
                      Bytes used      =       2550

  F=
    +theta(-2+N)*(
      +den(-1+N)*acc(23/8+ep^-2+1/3*ep^-1+4/3*z3)
      +den(-1+N)^2*acc(-7/2-ep^-1)
      +den(-1+N)^3*acc(3)
      +den(-1+N)^3*S(R(1),-1+N)*acc(2/3)
      +den(-1+N)^2*S(R(1),-1+N)*acc(-31/18)
      +den(-1+N)*S(R(1),-1+N)*acc(37/36+ep^-1)
      +den(-1+N)*S(R(1,1),-1+N)*acc(1)
      +den(-1+N)*S(R(2),-1+N)*acc(-29/18)
      +den(-1+N)*S(R(3),-1+N)*acc(-2/3) )
    +theta(-1+N)*(
      +den(N)*acc(-17/18-ep^-2-1/3*ep^-1+16/3*z3)
      +den(N)^2*acc(-73/18+ep^-1)
      +den(N)^3*acc(11/3)
      +den(N)^3*S(R(1),N)*acc(8/3)
      +den(N)^2*S(R(1),N)*acc(-29/9)
      +den(N)*S(R(1),N)*acc(23/9-ep^-1)
      +den(N)*S(R(1,1),N)*acc(-1)
      +den(N)*S(R(2),N)*acc(5/9)
      +den(N)*S(R(3),N)*acc(-8/3) )
    +theta(N)*(
      +den(1+N)*acc(4*z3)
      +den(1+N)^2*acc(8*z3)
      +den(1+N)^4*S(R(1),1+N)*acc(4)
      +den(1+N)^3*S(R(1),1+N)*acc(2/3)
      +den(1+N)^2*S(R(2),1+N)*acc(-2/3)
      +den(1+N)^2*S(R(3),1+N)*acc(-4)
      +den(1+N)*S(R(1),1+N)*acc(-8*z3)
      +den(1+N)*S(R(1,2),1+N)*acc(2/3)
      +den(1+N)*S(R(1,3),1+N)*acc(4)
      +den(1+N)*S(R(2,1),1+N)*acc(-2/3)
      +den(1+N)*S(R(3,1),1+N)*acc(-4) )
    +delta(-1+N)*(
      +acc(1001/648+4/3*ep^-3-2/9*ep^-2
                         -13/108*ep^-1-1/3*z3)
      )
    +delta(N)*(
      +acc(-1087/81-1/3*ep^-3-1/9*ep^-2
                         -97/54*ep^-1+20/3*z3)
      )
    +LA22(1,1+N,1,8,1,1,1,1,0,N,0,3)*acc(1);
\end{verbatim}
}

\section{Some extra remarks}

The above features are released in version 3.1 of FORM at its regular 
address http://www.nikhef.nl/$\sim$form.

Of course there are still features that FORM does not have and would be 
much appreciated. One would be proper GCD and factorization algorithms. 
This would make it much easier to solve sets of equations.
These are anticipated, but lack of manpower is the main problem.

It seems that the inherent speed of FORM comes from its internal 
data representation. The locality of its operations seems to be less 
important in this matter than was previously believed. This plays mainly a 
role when expressions are so big that they reside on disk. But even 
in that case a good use of the .sort instructions helps.

Currently a study is under way to see whether FORM can be made into an open 
source project. This would need a considerable amount of manpower, because 
the sources may have to be reprogrammed, several levels of documentation 
will have to be made and a number of additions will have to made.

\end{document}